\documentclass[%
 reprint,
 amsmath,amssymb,
 aps,
 prb,
 floatfix,
]{revtex4-1}

\usepackage{graphicx}% Include figure files
\usepackage{dcolumn}% Align table columns on decimal point
\usepackage{bm}% bold math
\begin{document}

\preprint{APS/123-QED}

\title{Variational Excitations in Real Solids: \\
       Optical Gaps
       and Insights into
       Many-Body Perturbation Theory}

\author{Luning Zhao$^1$}

\author{Eric Neuscamman$^{1,2,}$}%
\email{eneuscamman@berkeley.edu}

\affiliation{
${}^1$Department of Chemistry, University of California, Berkeley, CA, 94720, USA \\
${}^2$Chemical Sciences Division, Lawrence Berkeley National Laboratory, Berkeley, CA, 94720, USA
}

\date{\today}

\begin{abstract}
We present an approach to studying optical band gaps in real solids in which quantum Monte Carlo methods allow for the application of a rigorous variational principle to both ground and excited state wave functions.
In tests that include small, medium, and large band gap materials, optical gaps are predicted with a mean-absolute-deviation of 3.5\% against experiment, less than half the equivalent errors for typical many-body perturbation theories.
The approach is designed to be insensitive to the choice of density functional, a property we exploit in order to provide insight into how far different functionals are from satisfying the assumptions of many body perturbation theory.
We explore this question most deeply in the challenging case of ZnO, where we show that although many commonly used functionals have shortcomings, there does exist a one particle basis in which perturbation theory's zeroth order picture is sound.
Insights of this nature should be useful in guiding the future application and improvement of these widely used techniques.
\end{abstract}


\maketitle

The quantitative study of electronic excitations in solids remains a central topic in condensed matter theory due to their importance the spectroscopic characterization of materials and in technological applications such as light harvesting.
For many semiconductors, approaches based on many-body perturbation theory (MBPT) in the form of $GW$ \cite{Hedin1965} and Bethe-Salpeter equation \cite{Strinati1984} (BSE)  methods have been particularly successful \cite{onida2002electronic} and these and related methods remain a highly fruitful topic of research.
\cite{Hung2016,Kotliar2016,Kotliar2017,Neaton2017,Louie2017,Saito2017,Schilfgaarde2018,Chatterjee2018}
However, there remain many materials of great technological interest, especially within the transition metal oxides, whose low-energy excitations are poorly described by density functional theory (DFT) and MBPT.

Although MBPT does not in principle need to rely on input from DFT,
some of its most widely used practical incarnations (e.g.\ $G_0W_0$) assume 
%A key aspect of MBPT is the assumption of
a zeroth order picture in which
electronic excitations are simple particle-hole transitions between the
one-particle eigenstates of Kohn-Sham DFT with transition energies given
by differences between these Kohn-Sham orbitals' energies.
In this picture, the lowest excited state corresponds to a single open-shell Slater determinant in which one electron has been promoted from the valence band maximum (VBM) orbital to the conduction band minimum (CBM) orbital.
Although the DFT orbital energy difference is known to underestimate the corresponding band gap, \cite{onida2002electronic,Perdew2009} this zeroth order picture is nonetheless quite close to reality when solids like C diamond and Si are treated with standard LDA \cite{Perdew1981} or GGA \cite{Perdew1996} density functionals.
In these situations, the DFT orbitals closely resemble the excited electron and hole states and the orbital energy differences, although not perfect, are close enough to reality that MBPT variants that
perturb around them can be quite accurate. \cite{Shishkin2007}
The story can be strikingly different when a solid/functional pairing produces one-particle states that differ significantly from the true electron and hole states and/or the orbital energy differences stray too far from reality.
The success of hybrid functionals \cite{Adamo1999,Scuseria2003} in improving gap predictions in areas where pure functionals perform poorly \cite{Pela2015,Manousakis2015} implies that one or both of these issues can be sensitive to the fraction of exact exchange.
It is therefore not surprising that the reliability of MBPT can be strongly dependent on the choice of functional and what degree of self-consistency is sought in the $GW$ equations.
Were it possible to inspect the properties of the true excitonic wave function in challenging solids, one could hope to gain insight into why certain density functionals satisfy MBPT's assumptions better than others, and make the modeling of difficult materials' spectra substantially more predictive.

In this Letter, we present a variational formalism that enables accurate and systematically improvable predictions of a material's lowest excited state wave function and the corresponding optical gap, which can be used as a standalone predictive tool and as a window into the relationship between density functionals and the assumed zeroth order picture of MBPT.
Our approach combines recent advances in excited state variational principles
\cite{Zhao2016,Pj2017,Shea2017,ye2017sigma}
with a wave function ansatz suitable for both the ground and the VBM$\rightarrow$CBM state.
Crucially, the ansatz can describe both nontrivial BSE-like superpositions of particle-hole excitations and the dynamic polarizations of the electron cloud found in the vicinity of an exciton.
We stress that this approach employs energy differences between neutral states and thus probes optical gaps, and so exciton binding energies (EBE) must be considered when comparing to fundamental gaps.
Gap comparisons aside, the fact that the method yields an explicit wave function for the VBM$\rightarrow$CBM excitation allows us to directly inspect how well
a given density functional satisfies MBPT's zeroth order picture and thus how likely it is that accurate predictions will result.

Just as the energy $E=\langle\Psi|H|\Psi\rangle/\langle\Psi|\Psi\rangle$ can be minimized to find a ``variationally best'' ground state within a given ansatz, we find our excited state by minimizing
\begin{equation}
\Omega\left(\omega, \Psi\right) = \frac{\left<\Psi|\omega-H|\Psi\right>}{\left<\Psi|(\omega-H)^2|\Psi\right>}
=\frac{\omega-E}{(\omega-E)^2+{\sigma}^2}
\label{eqn:var}
\end{equation}
whose global minimum is not the ground state but the $H$-eigenstate with energy immediately above the chosen value $\omega$, \cite{Zhao2016} which we place within the band gap to target the first excited state and thus predict the optical gap.
To mitigate the difficulty of dealing with the $H^2$ term, our QMCPACK \cite{KimQMCPACK2018} implementation evaluates $\Omega$ via variational Monte Carlo (VMC) \cite{Umrigar:2015:qmc,Zhao2016} and minimizes it using the linear method. \cite{Zhao2016,Nightingale:2001:linear_method,UmrTouFilSorHen-PRL-07}
For a more detailed discussion of how this approach is kept size extensive \cite{Shea2017} and balanced between states, \cite{Pj2017}
as well as computational details and the addressing of finite size effects,
we refer the reader to the Supplemental Material (SM)
at the end of this document.

We pair this variational approach with a multi-Slater Jastrow \cite{ClarkTable2011,MoralesMSJ2012} ansatz
\begin{equation}
\label{eqn:msj}
\Psi(\vec{r}) = e^{U(\vec{r})} \sum_I C_I \Phi_I(\vec{r})
\end{equation}
where $U(\vec{r})$ is a correlation factor \cite{KimQMCPACK2018}
\begin{equation}
\label{eqn:jas}
U(\vec{r})=\sum_{ip}{V_p(r_{ip})}+\sum_{i<j}{W(r_{ij})}
\end{equation}
detailed in the SM.
This configuration interaction (CI) of the Slater determinants $\Phi_I$ is used to accommodate the basic structure of each state and account for state-specific polarization effects.
For the ground state, we include the closed shell Kohn-Sham determinant for the basic ground state structure plus all single particle-hole excitations, which represent the leading order terms in a Taylor expansion of the orbital rotation that would transform the Kohn-Sham determinant into whichever determinant minimizes $\Omega$ in the presence of the correlation factor.
For the excited state, we would like to include all single particle-hole excitations as in the BSE approach
as well as the closed shell determinant and all double particle-hole excitations.
This would again allow us to capture the leading order effects of an orbital rotation \cite{HeadGordon1994,Subotnik2011} that would in this case accommodate repolarizations of the electron cloud in the vicinity of the exciton.
However, as it is prohibitively expensive to include all double excitations in real materials, we approximate orbital relaxations by first minimizing $\Omega$ for singles and the closed shell term and then adding only those doubles that contain a singles component with coefficient larger than 0.1.

\begin{figure}[b]
\centering
%      trim={<left> <lower> <right> <upper>}
\includegraphics[clip,trim={0.5cm} {1.1cm} {0.1cm} {0.1cm}, width=8.5cm,angle=270]{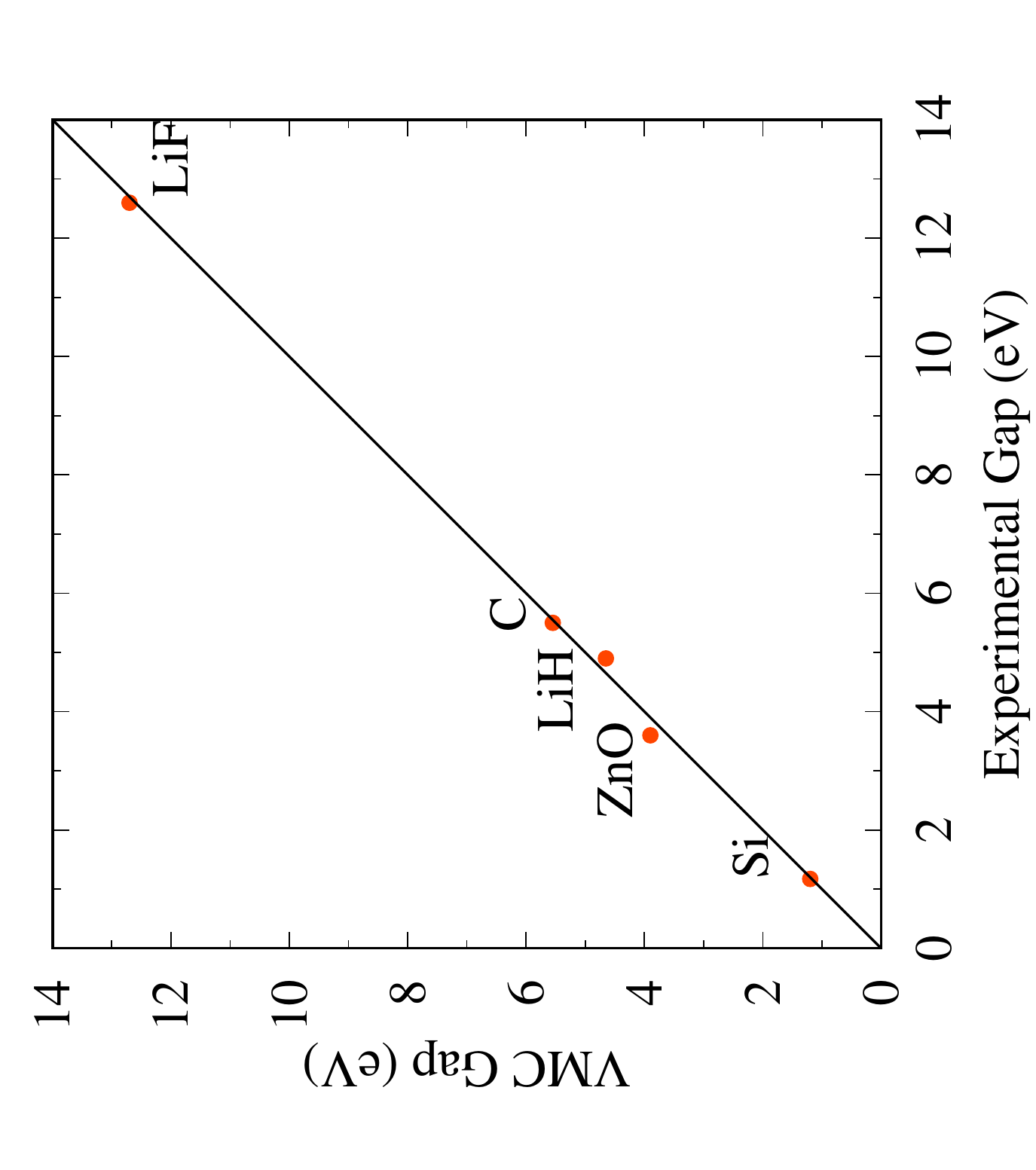}
\caption{VMC-CISD optical gap predictions plotted against experimental results.
         See Table \ref{tab:gaps} for more details.
        }
\label{fig:money}
\end{figure}

\begin{figure}[b]
\centering
\includegraphics[width=8.5cm,angle=0]{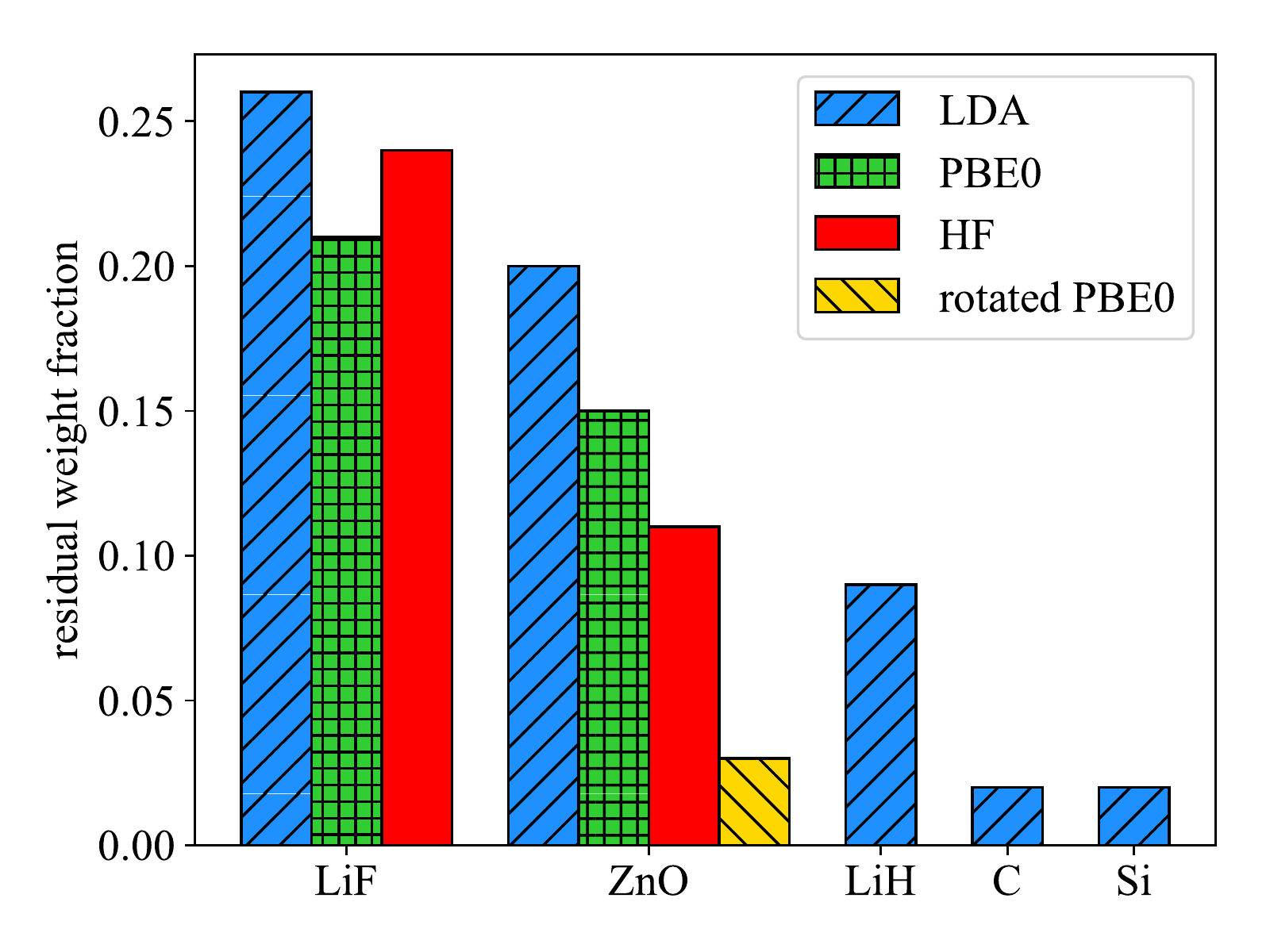}
\caption{Here we investigate the appropriateness of
         various one-particle orbital sets for MBPT by plotting
         VMC-CISD's residual weight fraction,
         which we define as the 
         sum of squared CI coefficients on all configurations
         other than the primary VBM$\rightarrow$CBM transition
         when working in a particular orbital basis.
         In cases where degeneracy in the VBM leads to multiple
         equal-energy VBM$\rightarrow$CBM configurations,
         the sum excludes all such configurations.
        }
\label{fig:ci}
\end{figure}

As seen in Figure \ref{fig:money} and Tables \ref{tab:gaps} and \ref{tab:zno}, the approach in which we include both singly and doubly excited configurations in the excited state (VMC-CISD) is quite effective for predicting optical gaps in small (Si), medium (C, LiH, ZnO), and large (LiF) band gap materials.
Its mean-absolute-deviation (MAD) from experimental values across these five systems is just 3.5\%, compared to MADs more than twice this large for the optical gaps obtained by subtracting the known exciton binding energies from $G_0W_0$ and self-consistent $GW$ gaps.
Of course, MBPT is highly effective in Si, C, and LiH, and so we expect that in these cases the zeroth order DFT wave function is sound.
The analysis in Figure \ref{fig:ci} confirms this expectation by showing that over 90\% of the VMC-CISD wave function is accounted for by LDA's VBM$\rightarrow$CBM transition.
Thus, in these three cases, LDA provides good zeroth order wave functions and we can confirm that the accuracy of MBPT derives from the appropriateness of its approximation.

\begin{table}[t]
\caption{Band gaps in eV.
         The quasiparticle gaps of DFT and the $GW$ methods should be reduced by
         the EBE when
         comparing to the VMC and experimental optical gaps.
         \label{tab:gaps}
}
\begin{tabular}{l r @{.} l r @{.} l r @{.} l r @{.} l}
\hline\hline
  &
\multicolumn{2}{ c }{ \hspace{0mm} C \hspace{0mm} } &
\multicolumn{2}{ c }{ \hspace{0mm} Si \hspace{0mm} } &
\multicolumn{2}{ c }{ \hspace{0mm} LiH \hspace{0mm} } &
\multicolumn{2}{ c }{ \hspace{0mm} LiF \hspace{0mm} } \\
\hline
LDA & 3&93 & 0&47 & 2&68 & 8&60 \\
$G_0W_0$ & 5&50\cite{Shishkin2007} & 1&12\cite{Shishkin2007} & 4&64\cite{Pickett2003} & 13&27\cite{Shishkin2007} \\
$GW$ & 5&99 \cite{Shishkin2007} & 1&28 \cite{Shishkin2007} & 4&75 \cite{Setten2007} & 15&10 \cite{Shishkin2007} \\
VMC-CIS & 5&68(6) & 1&41(6) & 5&01(6) & 14&6(1) \\
VMC-CISD & 5&55(6) & 1&20(6) & 4&65(6) & 12&7(1) \\
Experiment & 5&50 \cite{Chen1991} & 1&17 \cite{Chen1991} & 4&90 \cite{Plekhanov1976} & 12&6 \cite{Walker1967} \\
EBE & 0&07 \cite{Rossler2002} & 0&015 \cite{Wu2013} & 0&1 \cite{KINK1987138} & 1&6 \cite{Walker1967} \\
\hline\hline
\vspace{2mm}
\end{tabular}
\caption{ZnO band gaps and EBE in eV. \label{tab:zno}}
\begin{tabular}{l r @{.} l}
\hline\hline
LDA            & 0&83 \\
PBE0           & 3&03 \cite{Fuchs2007} \\
$G_0W_0$-LDA   & 2&14 \cite{Shishkin2007} \\
$GW$-LDA       & 3&20 \cite{Shishkin2007} \\
$G_0W_0$-PBE0  & 3&24 \cite{Fuchs2007} \\
VMC-CIS(LDA)   & 3&9(2) \\
VMC-CIS(PBE0)  & 4&6(2) \\
VMC-CISD(LDA)  & 3&9(2) \\
VMC-CISD(PBE0) & 3&9(2) \\
Experiment     & 3&6  \cite{Lauck2006} \\
EBE            & 0&06 \cite{Pulci2010} \\
\hline\hline
\end{tabular}
\end{table}

The story is quite different in LiF and ZnO, where Figure \ref{fig:ci} reveals that LDA's zeroth order picture accounts for less than 80\% of the high-level wave function.
At a minimum, this implies that LDA's VBM and CBM orbitals are not the correct shape for the real exciton's particle and hole, a point we will return to in our discussion of ZnO.
Figure \ref{fig:ci} also reveals that in these two systems, the fraction of exact exchange can have a significant effect on how closely DFT's zeroth order wave function matches the VMC prediction.
Although there are also the orbital energies to consider (see ZnO discussion below), these findings help explain why MBPT can be so sensitive to the choices made in its practical application. \cite{Louie2010,Friedrich2011,Berger2012}
Work by Sommer et al \cite{Pollmann2012} reveals that these issues can carry over to the BSE approach, which fails to provide a satisfactory correction to GW in LiF, although
vertex-corrected solutions to Hedin's equations can help in that case.
\cite{kutepov2016electronic,kutepov2017self}
Note that these issues do not necessarily imply a failure of one-particle theory in these systems, as there may exist a one particle basis in which the true exciton really does look like the simple VBM$\rightarrow$CBM transition.
Indeed, in ZnO, to which we will now turn our attention, we will provide an analysis showing that such a basis does indeed exist.
Thus, while Figure \ref{fig:ci} makes plain that commonly used density functionals struggle to meet the needs of MBPT in both ZnO and LiF, the insights gleaned from systematically improvable wave function methods should help resolve this difficulty in future. 

\begin{figure}[b]
\centering
\includegraphics[width=8.5cm,angle=0]{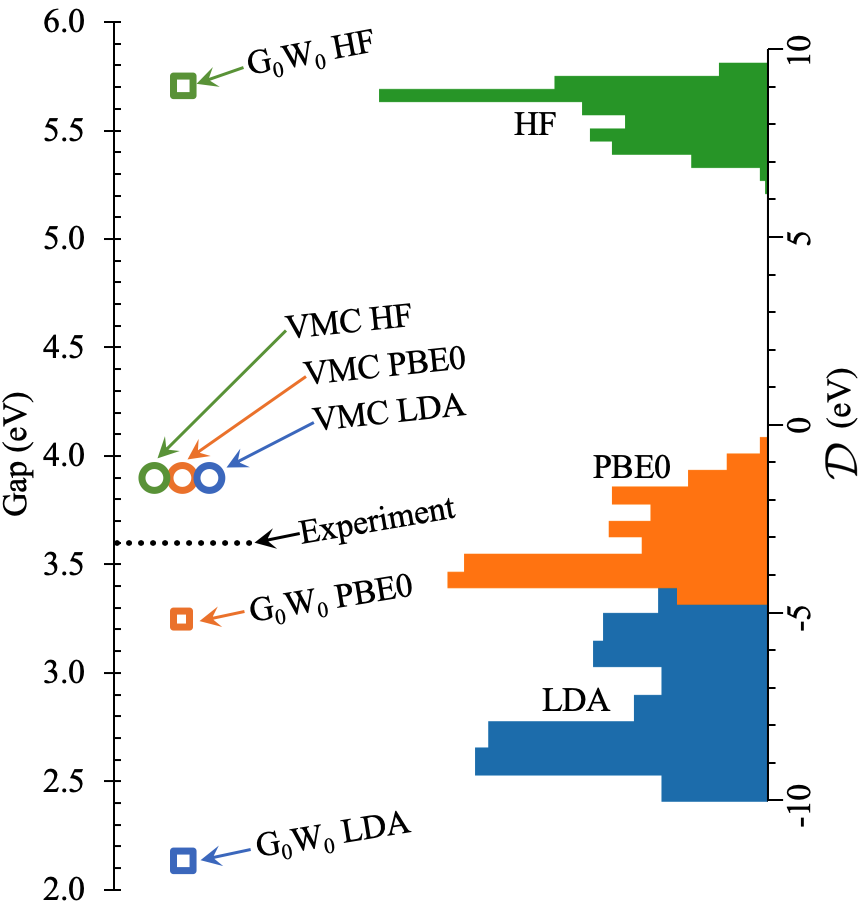}
\caption{Optical gap and single-particle transition energy data for ZnO.
         On the left, we compare G$_0$W$_0$ fundamental gaps using
         one-particle starting points that employ different fractions
         of exact exchange with our VMC-CISD optical
         gaps based on the same starting points.
         For the various $i\rightarrow a$ transitions,
         we plot on the right histograms of the differences
         $\mathcal{D}_{ia}=\Delta^{\mathrm{DFT}}_{ia}-\Delta^{\mathrm{VMC}}_{ia}$
         between the DFT estimates
         (i.e.\ the orbital energy differences
         $\Delta^{\mathrm{DFT}}_{ia}=\epsilon_a - \epsilon_i$)
         for the energy cost of promoting
         an electron from orbital $i$ to orbital $a$
         and the analogous quantities $\Delta^{\mathrm{VMC}}_{ia}$,
         which are the VMC energy differences between the
         $i\rightarrow a$ excited and the ground state
         Jastrow-modified Slater determinants.
         G$_0$W$_0$ data from Fuchs.
         \cite{Fuchs2007}
         Experimental result from Lauck. \cite{Lauck2006}
         }
\label{fig:zno_gaps}
\end{figure}

ZnO represents a particularly difficult case for MBPT, especially when considering its low-order and highly-efficient $G_0W_0$ variant.  \cite{Shishkin2007,Louie2010}
The left hand side of Figure \ref{fig:zno_gaps} makes clear that the accuracy of this low order perturbative treatment is highly sensitive to the inclusion of exact exchange.
In contrast, we see that the VMC-CISD results are insensitive to whether we employ the LDA, PBE0, or even the Hartree Fock (HF) one-particle basis sets.
The reasons for this success are two-fold.
First, the wave function was designed so as to be able to approximate an orbital rotation in order to counteract shortcomings in the starting DFT orbitals.
Indeed, if we remove this ability by removing the doubles excitations from the excited state and the singles from the ground state, the resulting VMC-CIS results are more sensitive and less accurate overall, as seen in Tables \ref{tab:gaps} and \ref{tab:zno}.
Second, VMC takes the issue of the DFT orbital energies off the table entirely, as it directly evaluates the energy expectation value of its wave function using the full ab initio Hamiltonian so that the only dependence on DFT is via the shapes of the one-particle orbitals.

Using our DFT-insensitive VMC methodology as a guide, one can investigate how commonly used density functionals' zeroth order pictures deviate from reality in ZnO and whether it is even possible to construct a one particle picture upon which MBPT should be reliable in this material.
First, we stress that although Figure \ref{fig:ci} revealed that $G_0W_0$'s sensitivity to exact exchange is likely due in part to the varying quality of the zeroth order wave functions, the right hand side of Figure \ref{fig:zno_gaps} emphasizes the importance of the zeroth order transition energies and how they are also quite sensitive to exact exchange.
By considering zeroth order wave functions and transition energies together, we gain an appreciation for how challenging this system is for density functional theory.
Indeed, HF theory with its 100\% exact exchange gives better orbitals for the purpose of describing the first excited state, but its transition energies are grossly too high, whereas PBE0 has better transition energies but worse orbitals.
Among the three options of LDA, PBE0, and HF, PBE0 clearly makes for the best compromise between wave function and transition energy accuracies, but our results suggest that both its energetics and orbitals would be improved in ZnO with a higher fraction of exact exchange.

\begin{figure}[b]
\centering
%      trim={<left> <lower> <right> <upper>}
\includegraphics[clip,trim={3.55cm} {7.5cm} {12.7cm} {5.68cm}, width=8.5cm]{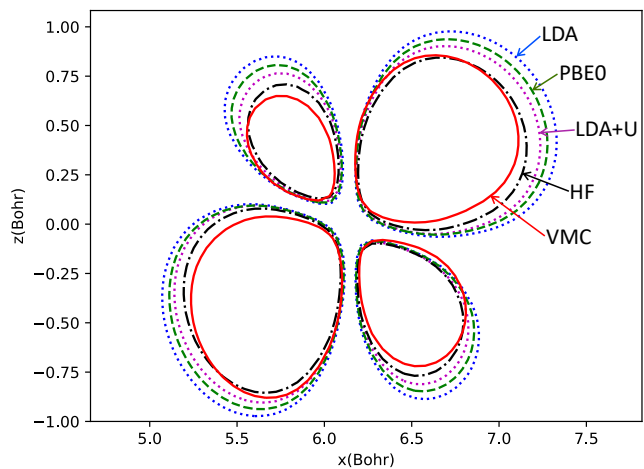}
\caption{A cut along ZnO's ($\bar{1}2\bar{1}0$) plane in which we investigate the
         lowest energy excitation's hole density in the vicinity of the Zn atom.
         For each method, we plot the contour along which the number of 
         holes per \AA$^3$ is equal to 1.2.
        }
\label{fig:hole}
\end{figure}

With an explicit high-level wave function in hand, we can ask highly detailed questions about the exciton, such as to what degree the O 2p and Zn 3d orbitals hybridize in the hole state.
Indeed, metal-oxide over-hybridization has been pointed out as a key deficiency in LDA and other pure functionals. \cite{Louie2010} 
We approach this question by performing a density matrix difference analysis \cite{head1995analysis}
in which the difference between the one-body density matrices of our VMC-CISD excited and ground state
is diagonalized.
As occurs for any excited state consisting of a single one-particle transition of the type assumed in MBPT's zeroth order picture, \cite{head1995analysis} the resulting eigenvalues are all close to zero (absolute values less than 0.1) except for one with a value near 1 and one with a value near -1.
The eigenvectors corresponding to these two large eigenvalues are the attachment and detachment orbitals, respectively, and represent the particle and hole orbitals that most closely represent the transition between a correlated many-body ground state and excited state.
By plotting the hole density from this detachment orbital in the vicinity of the Zn atom alongside the hole densities predicted by the VBM of different density functionals, Figure \ref{fig:hole} makes clear that, compared to our high-level VMC results, LDA does indeed include too much Zn character in the VBM through over-hybridization.
More surprisingly, we see that although LDA+U \cite{Cococcioni2014} with the U value used previously \cite{Louie2010} does decrease the degree of hybridization, our detachment density is even less hybridized, with LDA+U bringing us only about halfway in between the LDA and VMC extremes.
Another important point that the detachment orbital reveals is that some hybridization is definitely present, just not so much as common density functionals, even those specifically designed to address this issue, predict.

Although it is frustrating that current functionals face the various difficulties discussed above, the fact that the VMC density difference analysis strongly resembles a simple single-particle transition suggests that it should be possible to design a functional that delivers an excellent zeroth order starting point for MBPT.
To make this idea more concrete, we can test whether such an orbital basis exists by applying an orbital rotation to our wave function (starting with the optimized VMC-CISD state in the PBE0 orbitals) in order to minimize the residual weight fraction of the exciton.
As seen in Figure \ref{fig:ci}, this rotated PBE0 one-particle basis matches the assumptions of MBPT in ZnO almost as well as the LDA basis does for Si or diamond.
This finding also serves to reassure us that the error we do see in VMC-CISD's optical gap prediction (and its moderate disagreement with previous projector Monte Carlo
estimates \cite{ma2013excited,yu2015towards,Santana2015})
is most likely due to the imperfect nature of our finite size correction rather than to the appropriateness of our wave function approach, as it validates the assumption that the excitonic state is dominated by single particle-hole transitions with the doubles only contributing small corrections.
While a good one-particle basis is just a start (density functionals must also produce reasonable zeroth order transition energies) the insights we now have from VMC paint a bright picture for the prospects of increasing the accuracy and reliability of MBPT in cases like ZnO.

We have shown that an excited state variational principle can be combined with simple, physically-motivated wave function approximations to evaluate optical band gaps in a way that is both insensitive to the DFT starting point and informative about the assumptions of MPBT.
Given the dominant role that MBPT plays in the theoretical interpretation of materials spectroscopy, a method that is able to improve its predictive power has the potential to be highly impactful.
Even in cases where exciton-induced repolarization effects are large and it is not possible to identify a density functional that yields a one-particle basis appropriate for describing both the ground and the low-lying conduction band states, the ability to provide variational predictions of band-edge energies, perhaps even in a k-point-by-k-point fashion, would create the possibility of developing first-principles-based scissors corrections for the BSE Hamiltonian, a practice that at present can be quite effective when performed empirically. \cite{Pulci2010,Bechstedt2009} 
In molecular excitations, variational excited states \cite{Zhao2016,Pj2017,Shea2017,ye2017sigma}
and MBPT \cite{Azarias2017}
have so far been explored separately, but the same potential for strong synergies is present.
In both molecules and solids, our approach also provides a reasonably black-box route to producing high-quality nodal surfaces for excited states in diffusion Monte Carlo, which even with less sophisticated VMC preparations has already shown promise in evaluating band gaps.
\cite{Mitas2008,Mitas2010,Abbasnejad2012,Ertekin2013,Wagner2014,Santana2015}
The prospects for increased accuracy and scalability in this area are especially bright in light of recent progress in VMC methods for optimizing the one-particle basis \cite{Filippi2016,Filippi2017} and achieving compact representations of excited states, \cite{Nick2017,Nick2018} not to mention the rapid progress in selective CI methods that synergize strongly with multi-Slater VMC.
\cite{Knowles2015,Hoffmann2016,Evangelista2016,Tubman2016,Holmes2016,Sharma2017,Zimmerman2017,Ohtsuka2017}
With this wide range of promising connections, we look forward to further exploring the role that variational approaches can play in deciphering and designing molecular and materials spectra.

%\emph{Acknowledgement}
\begin{acknowledgments}
We thank Paul Kent and Jeffrey Neaton for helpful discussions.
This work was supported by the U.S. Department of Energy, Office of Science, Basic Energy Sciences, Materials Sciences and Engineering Division, as part of the Computational Materials Sciences Program and Center for Predictive Simulation of Functional Materials.
Calculations were performed using the Berkeley Research Computing Savio cluster.
%Input and output files for our calculations
%are available (DOI: 10.18126/M2K63Q)
%via the Materials Data Facility.
\end{acknowledgments}

% ========================================== BEGIN SUPP MAT ==========================================
\clearpage
{\Large \noindent \textbf{Supplemental Material}}

\hphantom{\small{ABC}}

\noindent 
%\section{Optimizing $\Omega$}
{\large Optimizing $\Omega$}

\hphantom{\small{ABC}}

Between linear method steps in the minimization of $\Omega$, we gradually
relax \cite{Shea2017} $\omega$
so that at convergence we achieve $\omega=E-\sigma$ and a result that is
equivalent to that of a size-extensive minimization \cite{FilippiEnVar2005}
of the energy variance $\sigma^2$.
Crucially, each linear method step seeks to minimize the function $\Omega$
in which $\omega$ is simply a constant, which guarantees that we target the
state with energy just above $\omega$.
This differs from a direct minimization of $\sigma^2$, which is not state-specific.
Note that we also use $\Omega$ to evaluate the ground state, as this is known \cite{Zhao2016} to provide a more balanced treatment for energy differences than optimizing one state with $\Omega$ and the other with $E$. 
To further improve balance so as to maximize cancellations of error, we also follow the recent approach \cite{Pj2017} of adjusting the flexibility of one of the wave functions in order to ensure that, as measured by $\sigma^2$, the ground and excited wave functions are of equal quality.
While this variance matching could be achieved by limiting the flexibility of either the ground or the excited state, we have done so in this study by withholding enough high energy singles from the ground state such that its variance matches that of the excited state.

\hphantom{\small{ABC}}

\noindent 
%\section{Computational details}
{\large Computational details}

\hphantom{\small{ABC}}

We have implemented our method within a development version of QMCPACK, \cite{KimQMCPACK2018} in which we have adapted the fast multi-Slater method \cite{ClarkTable2011,MoralesMSJ2012} to work with the cubic B-spline representation \cite{KimQMCPACK2018} of Kohn-Sham orbitals imported from {\sc Quantum ESPRESSO}. \cite{QE2009,QE2017}
We also customized the linear method optimizer to support complex numbers.
For the correlation factor $U$ in the main text, QMCPACK represents the one-dimensional functions $V$ and $W$ by 10-point cubic B-splines of the electron-nuclear ($r_{ip}$) and electron-electron ($r_{ij}$) distances. \cite{KimQMCPACK2018}
Note that independent spline parameters are used for each chemical element and for same- and opposite-spin electron pairs, and that the parameters are optimized alongside the configuration interaction coefficients during the minimization of $\Omega$.

To avoid the unnecessary simulation of low-energy core electrons, we used
Burkatzki-Filippi-Dolg (BFD) pseudopotentials \cite{Dolg2008} for Li, C, F, and Si, the norm-conserving pseudopotential of Shin et al \cite{Shin2017} for O, and the semi-core-included pseudopotential
of Krogel et al \cite{Krogel2016} for Zn.
All DFT calculations were performed with QUANTUM ESPRESSO 5.3.0 using a 350-Ry kinetic
energy cutoff and a 4$\times$4$\times$4 k-point grid.
All lattice constants were chosen based on experimental values,
with an fcc lattice structure used for LiH and LiF with lattice constants of 
7.716 and 7.625 Bohr, respectively.
The diamond cubic structure was used for both C diamond and Si with lattice constants of 6.740 
and 10.263 Bohr, respectively.
The wurtzite structure was used for ZnO with lattice constants set to $a=3.250$ Bohr and $c=5.207$ Bohr.
For ZnO, a 4-atom unit cell was used for DFT, while all other systems used a 2-atom unit cell for DFT.

\begin{figure}[b]
\centering
\includegraphics[scale=0.9,angle=0]{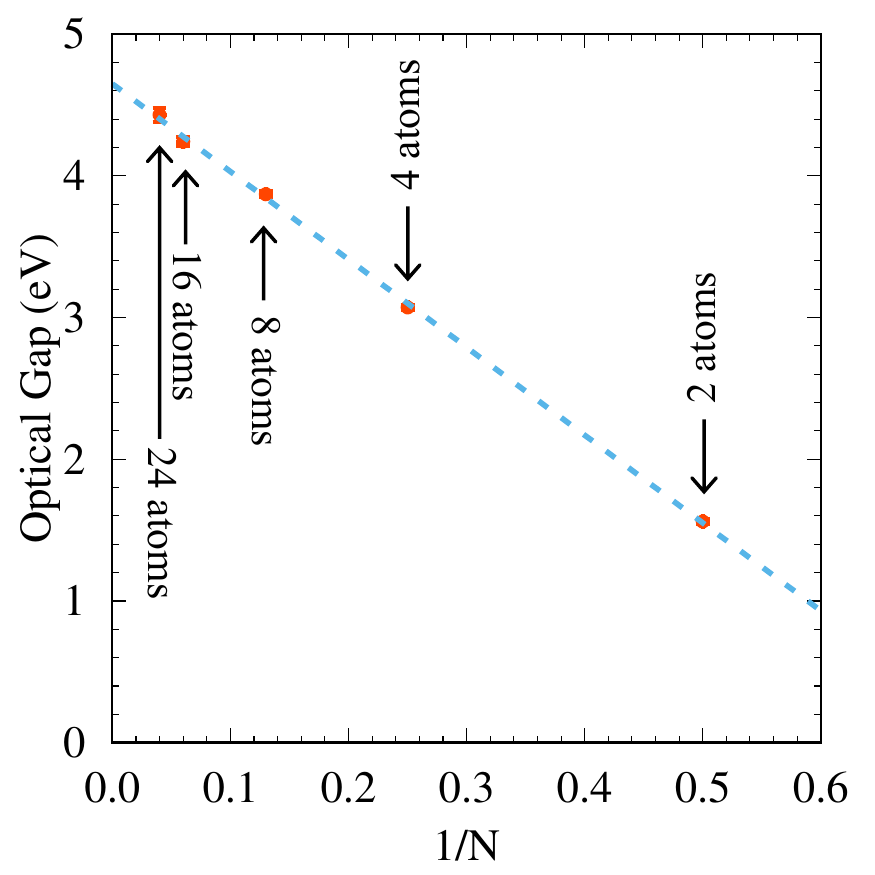}
\caption{Extrapolation to the bulk limit in LiH.  Points are our VMC data while the dashed line is a linear fit against the inverse of the number of atoms $N$ in the simulation cell.}
\label{fig:lih}
\end{figure}

VMC calculations for LiH and LiF were performed in simulation cells containing 2, 4, 8, 16, and 24 atoms,
after which $1/N$ extrapolations were used to predict the band gap in the bulk limit.
The same approach was used for C diamond and Si, but with 8, 16 and 24 atom simulations cells.
See Figure \ref{fig:lih} for an example extrapolation.

Due to the high cost of simulating the semi-core electrons of Zn, which was necessary to produce
physically reasonable results, we were limited by our current software implementation to a maximum
of 8 atoms in our simulation cell for ZnO,
which did not permit us to perform the same type of finite size correction as for the systems above.
Instead, we have derived a simple finite size correction based on previous
diffusion Monte Carlo (DMC) work \cite{Santana2015}
in which nodal surfaces for both the CBM and VBM were constructed using a simple single-Slater model.
The previous study reports results for a 48 atom simulation cell, and so we have performed the
equivalent single-Slater DMC calculations for our 8 atom cell and used the difference in the DMC gap
at these two cell sizes to provide an approximate finite size correction for our 8 atom VMC gap.
Note that this approach has no effect on our conclusions with regard to either the nature of the
first excited state under different density functionals or the insensitivity of our VMC gap predictions to
the choice of functional, as these properties are entirely determined within our 8 atom VMC evaluations.

% ========================================== END SUPP MAT ==========================================

\bibliography{main}

\end{document}